%% file: main.tex
\pdfoutput=1
\documentclass[9pt,letterpaper,twocolumn]{extarticle}

\usepackage{amssymb}
\usepackage[bookmarks=true,bookmarksopen=true]{hyperref}
\usepackage[version=3]{mhchem}

\usepackage{latextuning}
\usepackage{ACSPhotonics_mimicry}

\begin{document}
\input{cap.tex}
\input{abstract.tex}
\maketitle
\input{introduction.tex}
\input{results.tex}
\input{discussion.tex}
\input{methods.tex}
\input{suppdescr.tex}
\authorinformation
\input{acknowledgements.tex}
\input{main.bbl}

\input{supporting.tex}
\input{supporting.bbl}
\end{document}

%% file: cap.tex
\title{Fundamental limits to far-infrared lasing in Auger-suppressed HgCdTe quantum wells}


\author{Georgy Alymov,\hyperlink{alymov}{$^{\displaystyle *}$}\hyperlink{MIPT}{$^\dagger$} Vladimir Rumyantsev,\hyperlink{IPM_RAS}{$^\ddagger$}$^,$\hyperlink{UNN}{$^\S$}\hyperlink{ORCID}{\ORCIDicon} Sergey Morozov,\hyperlink{IPM_RAS}{$^\ddagger$} Vladimir Gavrilenko,\hyperlink{IPM_RAS}{$^\ddagger$} Vladimir Aleshkin,\hyperlink{IPM_RAS}{$^\ddagger$} and Dmitry Svintsov\hyperlink{MIPT}{$^\dagger$}
}

\affiliations{\hypertarget{MIPT}{$^\dagger$}Moscow Institute of Physics and Technology, Dolgoprudny 141700, Russia\\
\hypertarget{IPM_RAS}{$^\ddagger$}Institute for Physics of Microstructures of RAS, 603950 Nizhny Novgorod, Russia\\
\hypertarget{UNN}{$^\S$}Lobachevsky State University of Nizhny Novgorod, 603950 Nizhny Novgorod, Russia}

\email{alymov}{$^{\displaystyle *}$}{alymov@phystech.edu}

\ORCID{Vladimir Rumyantsev}{0000-0003-0739-2214}

%% file: abstract.tex
\tocfigure{figures/eye_candy}
\abstract{A challenge of bridging the terahertz gap with semiconductor lasers faces an inevitable problem of enhanced non-radiative Auger recombination with reduction of photon energy. We show that this problem can be mitigated in mercury-cadmium-telluride quantum wells (HgCdTe QWs) wherein the Auger process is suppressed due to formation of quasi-relativistic electron-hole dispersion imposing strong energy-momentum restrictions on recombining carriers. Such dispersion is formed upon interaction of topological states at the two QW interfaces. We characterize the lasing properties of HgCdTe QWs quantitatively by constructing a microscopic theory for recombination, absorption, and gain, and show the feasibility of lasing down to $\sim 50$ $\mu$m at liquid nitrogen temperature with threshold currents two orders of magnitude lower than in existing lasers. Our findings comply with recent experimental data on stimulated far-infrared emission from HgCdTe QWs and show the directions toward achievement of maximum possible lasing wavelength.}

%% file: introduction.tex
\begin{introduction}[S]emiconductor lasers operating in the far-infrared (FIR) and terahertz frequency ranges are highly demanded for biomedical imaging~\cite{THz-Imaging}, atmospheric spectroscopy~\cite{MIR_lasers_spectroscopy}, trace gas detection~\cite{FIR_trace_gas}, and disclosure of explosives and drugs~\cite{THz_drug_detection}. The main problem toward creation of such lasers is finding a material with suitably low band gap $E_g$ and simultaneously long non-radiative carrier lifetime. This lifetime is dramatically shortened upon gap shrinkage due to rapid electron-hole annihilation with energy-momentum transfer to another carrier, known as Auger recombination (AR). The AR rate typically demonstrates an activation-type behaviour $R_A \propto e^{-E_{\rm th}/kT}$, where $T$ is the temperature and $E_{\rm th}$ is the threshold energy being proportional to the gap~\cite{Abakumov-nonradiative}.

Activation-type behaviour of Auger process prohibits the operation of laser diodes based on narrow-gap materials at non-cryogenic temperatures~\cite{InSb_laser,Arias_HgTe_laser,Malin_MIR_lasers}. For this reason, quantum-cascade lasers (QCLs) become superior at 1--5 THz frequencies~\cite{Belkin_QCLs}, but they cannot operate in the 5--10 THz window~\cite{QCL_forbidden_region,QCL_28.3um_77K} where lattice absorption and phonon-assisted depopulation of lasing levels are strong~\cite{Belkin_QCL_limiting}. Still, laser diodes have much simpler design compared to QCLs, and if the problem of AR could be mitigated, they can beat QCLs in cost efficiency at most wavelengths. 

The Auger activation threshold $E_{\rm th}$ depends strongly on the energy-momentum relation for electrons and holes $\varepsilon_p$. Therefore, engineering the band structure by tuning strain~\cite{Auger_strain_engineering}, boundary profile of quantum wells~\cite{QW_boundary_AR_suppression} and dots~\cite{Efros_AR_suppression}, and their composition remained so far the most efficient way to suppress AR. At the same time, there exists a peculiar type of band dispersion where AR is prohibited by energy-momentum conservation laws. It resembles the electron-positron dispersion in vacuum~\cite{Dirac1930} and reads $\varepsilon_p^2 = p^2 v_0^2+E^2_g/4$, where $v_0$ is the band velocity. Impossibility of AR here stems from non-equal energies of initial ($\ge 3mv_0^2$) and final ($mv_0^2$) particles in the center-of-mass frame.

Finding a narrow-gap material with nearly perfect Dirac dispersion would resolve the problem of AR and FIR lasing. Previous expectations relied on lead salts~\cite{Dimmock_LeadTin,Emtage_AR_lead_tin} and graphene~\cite{Ryzhii_NDCGraphene}. In the former case, achievement of lasing is hampered mainly by technological problems (high residual doping). In the latter case, AR cannot be fundamentally switched off due to many-particle effects~\cite{Gierz2015,our}.

In this paper, we theoretically show that mercury-cadmium-telluride quantum wells (HgCdTe QWs) below the topological transition represent an ideal platform for AR suppression and FIR lasing. A quasi-relativistic electron-hole dispersion is realized therein as a result of hybridization between topological states at two adjacent \ce{Cd$_y$Hg$_{1-y}$Te/HgTe} interfaces~\cite{BHZ}. We argue that Auger suppression due to ``diracness'' of electron-hole bands has driven the recently achieved long-wavelength (up to $\lambda \sim 20$ $\mu$m) stimulated emission in narrow QWs~\cite{experiment-old,experiment,Experiment-OE}, and the lasing wavelength can further reach up to $\sim50$ $\mu$m at liquid nitrogen temperature. This possibility was overlooked during half-a-century history of CdHgTe photonics, as most laser studies focused either on wide wells~\cite{Bleuse_HgTeQWLaser,Roux_DFB_HgTe_laser} or wells with large Cd fraction~\cite{Meyer_HgTe_theory}. Through the detailed theoretical calculations, we show that optimal conditions for FIR lasing are realized in narrow wells of pure HgTe.

The paper is organized as follows. We first calculate the electron-hole spectra in HgCdTe quantum wells, elucidate the formation of Dirac-type dispersion at specific QW thicknesses, and demonstrate an enhancement of Auger threshold. We show that Auger suppression occurs only in narrow (non-topological wells), while above the topological transition the QW band structure enables nearly thresholdless Auger recombination~\cite{normal_vs_inverted}. We further proceed to detailed calculations of absorption, laser gain and threshold carrier densities for lasing in population-inverted QWs, and evidence the advantage of reduced inter-subband absorption in non-topological wells. Finally, we calculate the non-radiative recombination rates and threshold pumping intensities/currents for HgCdTe QWs and show their superiority with respect to lead salt lasers and QCLs at specific wavelengths. We also theoretically demonstrate HgCdTe laser operation in the 5-10 THz range which cannot be covered by GaAs lasers due to lattice absorption.
\end{introduction}

%% file: results.tex
\section{Results}
\subsection{Auger suppression in HgCdTe quantum wells}

\begin{fig}{band_edges}{figures/Band_edges}
 Solid curves: position of band edges in HgCdTe QWs of different thicknesses at lattice temperature of 4.2 K. Dashed curve: valence band position at the center of the Brillouin zone. Insets: bandstructures of normal-band (narrow) wells, wells close to the topological transition, and inverted-band (wide) wells. Vertical lines indicate the topological transition at $d_c = 6.3$~nm and the transition to indirect gap at $d_{\rm in} = 7.1$~nm.
\end{fig}

 Modeling of lasing and non-radiative recombination in \ce{Cd_{$y$}Hg_{$1-y$}Te/Hg_{$x$}Cd_{$1-x$}Te/Cd_{$y$}Hg_{$1-y$}Te} quantum wells (HgCdTe QWs) requires the knowledge of electron-hole spectra and wave functions. To calculate the latter, we have developed a bandstructure solver based on anisotropic eight-band Kane $\vec{k}\cdot\vec{p}$-model (see Methods) \cite{Kane_model}. To stay relevant to experimental results of~\Cite{experiment}, we restrict our consideration to QWs with $x=1.0$ and $y=0.7$ grown along the [013] direction, though the main features of spectrum and evolution of lasing properties are generic (see \hyperlink{discussion}{Discussion}). The bandstructure of HgCdTe QWs depends on the lattice temperature $T_{\rm lattice}$ and, to avoid confusion, we assume $T_{\rm lattice} = 4.2$~K throughout this section (except in \autoref{fig:current-frequency}b). The results for $T_{\rm lattice} = 77$~K and 300 K are shown in \suppref{fig:n_and_t}. 

 Upon increasing the well thickness, it undergoes a topological transition from a normal semiconductor (CdTe-like) to a topological insulator (HgTe-like) (\autoref{fig:band_edges}). At critical thickness $d_c = 6.3$~nm, two-dimensional states in a HgCdTe QW are gapless with the conduction and valence bands forming a Dirac cone~\cite{BHZ}, as shown in \autoref{fig:band_edges}. At a slightly larger thickness, $d_{\rm in} = 7.1$~nm, the spectrum is transformed from direct- to indirect-gap, with valence band edge forming a circle in momentum space~\cite{anisotropynote}. In immediate vicinity of critical thickness, HgCdTe QWs are narrow-gapped with Dirac-type electron-hole dispersion.
 
 \bibnotetext[anisotropynote]{Actually, the side maximum of the valence band is circular only in the isotropic approximation. When the band anisotropy present in HgCdTe QWs is taken into account (which we did, see \suppref{fig:all-in-one}), the valence band edge consists of only two points in the momentum space. However, this still results in a zero-threshold Auger process along at least some directions in the $k$-space, and in elevated density of states near the valence band edge. In view of the above, we will call this maximum `circular' for simplicity.}

\begin{fig}{Auger_thresholds}{figures/Auger_thresholds}
 Solid curve: Auger threshold energy $E_{\rm{th}}$ in units of bandgap $E_g$ in HgCdTe QWs of different thicknesses at lattice temperature of 4.2 K. Dashed curve: Auger threshold in the parabolic-band approximation. Inset: threshold energies in absolute units. The behavior of threshold energies is explained schematically by the difference in band structures and threshold Auger processes for normal, nearly-critical and topological wells (from left to right). Separate electron and hole contributions to $E_{\rm th}$ are shown in \suppref{fig:e-h_thresholds}.
\end{fig}

The degree of AR suppression due to peculiar electron-hole spectrum can be described by a single quantity, the threshold energy $E_{\rm{th}}$. Here, we define it as the net kinetic energy of three involved particles: two electrons and one hole in the CCCH process or two holes and one electron in the CHHH process, whichever is smaller. This energy appears in the Boltzmann exponent for Auger coefficient for nondegenerate carriers.

The calculated threshold energy vs QW thickness is shown in \autoref{fig:Auger_thresholds} with solid line. The dashed line shows the same quantity in the parabolic-band approximation, where $E_{\rm{th}}/E_g = \min\{m_e, m_h\}/(m_e + m_h)$ and peaks at half of the gap for mirror-symmetric spectra. We readily observe that true energy threshold strongly exceeds its value for parabolic-band QWs due to formation of Dirac-like dispersion near the critical thickness. The threshold normalized to the gap value, $E_{\rm{th}}/E_g$, diverges near $d_c$. However, $E_{\rm{th}}$ itself tends to zero at the threshold. The real increase of threshold energy due to ``diracness'' reaches up to 12 meV and leads to sixfold extra AR suppression at liquid nitrogen temperature.

Above the critical thickness, the threshold energy increases following the trend of gap. However, at the direct-to-indirect transition, a recombination channel involving two holes near the circular extremum is switched on (shown in the rightmost bandstructure in \autoref{fig:Auger_thresholds})~\cite{normal_vs_inverted}, which has vanishing energy threshold. Due to this process, stimulated interband emission and lasing have never been observed in wide HgCdTe quantum wells~\cite{narrow_vs_wide}.

\begin{fig}{concentrations}{figures/Concentrations}
 Threshold carrier concentrations required for achieving optical gain in HgCdTe QWs of different thicknesses. Dashed curves correspond to threshold limited by pure inter-subband absorption ($\gamma = 0$), solid lines---to sum of Drude and inter-subband absorption with $\gamma = 1$~meV. Lattice temperature is 4.2 K, electron temperature is 300 K (red curves) or 77 K (blue curves). Results for $T_{\rm lattice} = T_e$ are shown in \suppref{fig:n_and_t}a.
\end{fig}

\subsection{Threshold of stimulated emission}

Though threshold energy is a simple and illustrative parameter governing the strength of AR, more practical figures of merit for interband lasers are the recombination rate $R$ and recombination time $\tau_r$ \emph{at the threshold of stimulated emission} (SE). Threshold carrier density for SE is determined by the balance of interband gain and intraband/intersubband absorption. We have calculated these quantities using the Fermi golden rule with electron-photon matrix elements evaluated on wave functions from our bandstructure solver. From now on, electron distributions are assumed to obey Fermi statistics within each band, with distinct quasi-Fermi levels and equal temperatures. 

The intersubband absorption can not be avoided by reduction of temperature and increasing sample quality, and therefore sets the true fundamental limit to the threshold density. Its value is shown in \autoref{fig:concentrations} with the dashed line. The threshold density first goes down with reduction of the gap, reaches a minimum slightly before $d_c$, and then goes up until the indirect-gap region where it remains almost constant. The initial decrease in carrier density is associated with lighter hole mass (and smaller density of states) in narrow-gap QWs. The subsequent increase in threshold density in wider wells is associated with small subband spacing (see \suppref{fig:all-in-one}) and multiple channels for inter-subband absorption (\suppref{fig:optical_conductivity}). Further on, the hole density of states becomes singular at the direct-indirect transition, which increases the threshold hole density at the desired quasi-Fermi level.

The magnitude of Drude absorption is governed by momentum relaxation rate $\gamma = \hbar/\tau_p$ which we have taken to be 1 meV. This corresponds to carrier mobility $\sim 6 \times 10^4$~cm$^2$/(V~s) for electron mass of $0.02 m_0$, a value readily achieved experimentally~\cite{Kvon_mobility}. As seen from \autoref{fig:concentrations}, Drude losses affect threshold density only in the vicinity of zero-gap state (where lasing frequency is small) and in wide wells (where the carrier density is high). At the same time, they have almost no effect in narrow QWs where Auger suppression is most pronounced.

\begin{fig}{recombination_times}{figures/Recombination_times}
 Auger (solid curves), phonon (dashed curves), and radiative (dotted curves) recombination times in HgCdTe QWs of different thicknesses at threshold carrier density. Lattice temperature is 4.2 K, electron temperature is 300 K (red curves) or 77 K (blue curves). Results for $T_{\rm lattice} = T_e$ are shown in \suppref{fig:n_and_t}b. Error bars show the standard deviations of Monte Carlo integration.
\end{fig}

\begin{widefig}{current-frequency}{figures/Current-frequency}
 Threshold currents vs lasing frequencies of \ce{Cd_{0.7}Hg_{0.3}Te/HgTe/Cd_{0.7}Hg_{0.3}Te} QWs of various thicknesses (threshold current and lasing frequency vs thickness are shown in \suppref{fig:currents_and_gaps}). Threshold intensities of optical pumping are shown on the right axes, assuming 0.4\% absorption of a 2.15 $\mu$m pump, as in \Cite{experiment}. (a) Lattice is held at 4.2 K, (b) lattice temperature equals electron temperature. Electron temperature is 300 K (red curves) or 77 K (blue curves). Insets in panel (a): threshold Auger processes in normal-band wells (with low threshold currents) and inverted-band wells (with high threshold currents). Green circles: experimental data on stimulated emission from photoexcited QWs~\cite{experiment} (unknown carrier temperature). Triangles and squares: threshold currents of existing quantum cascade lasers and interband cascade lasers at 300 K (red) or 77 K (blue) (Refs.~\citenum{ICL_3.67um_300K,QCL_5.1um_300K,ICL_6um_300K,ICL_7um_300K,QCL_7.77um_300K,ICL_10.4um_77K,QCL_15.1um_300K,QCL_17.8um_77K,QCL_24.5um_77K} and \citenum{QCL_28.3um_77K}, ordered by increasing wavelength). Circles: threshold currents of \ce{Pb_{$1-x$}Sn_{$x$}Se} diffusion diode lasers \cite{Lead_Salt_LaserCharacteristics} and \ce{Pb_{$1-x$}Sn_{$x$}Se_{$1-y$}Te_{$y$}} double heterostructure diode lasers~\cite{LeadSalt_DoubleHeterostructure,leadsaltnote}. 
 Reststrahlen bands of HgCdTe and GaAs are colored red and green.
\end{widefig}

\bibnotetext[leadsaltnote]{In \Cite{LeadSalt_DoubleHeterostructure}, the lasing wavelength is $\lambda = 20 \mu$m at 20 K. We assume it decreases to 15 $\mu$m at 77 K based on the known temperature dependence of $\lambda$ in \ce{Pb_{$1-x$}Sn_{$x$}Se}~\cite{Lead_Salt_LaserCharacteristics}.}

Already a small carrier density for SE and high AR threshold make narrow HgCdTe QWs ($d < d_c$) superior to wide ones ($d > d_c$) in terms of lasing properties. This suggestion is further supported by full calculations of recombination times due to AR, interband phonon emission, and spontaneous photon emission, shown in \autoref{fig:recombination_times}. The AR times rapidly go down until the critical thickness and remain almost constant afterwards, at the temperature-independent level of $\sim 0.3$--0.5 ps. A similar reduction of AR time has previously been predicted for inverted-band HgCdTe superlattices~\cite{normal_vs_inverted}.

Recombination through the emission of optical phonons is generally slower, as shown in \autoref{fig:recombination_times} with dashed line. This occurs primarily due to low optical phonon energy~\cite{phonon_params} $\hbar \omega^{\rm max}_{\rm op}=19$~meV. As a result, away from the critical thickness, only multiphonon recombination is possible.

Radiative recombination due to spontaneous photon emission dominates carrier kinetics in photoexcited mid-infrared wells at cryogenic temperatures~\cite{HgCdTe_Recombination}, but becomes irrelevant in far-infrared region or at higher temperatures. Note that it does not mean the infeasibility of far-infrared HgCdTe lasers, because it is \emph{stimulated} emission which drives laser operation, whose rate can be arbitrarily higher than the spontaneous emission rate.

\subsection{Prospects for CdHgTe interband infrared lasers}

Recombination suppression in HgCdTe QWs can result in improved figures of merit for QW-based lasers. These are reduced threshold intensities for optically pumped systems and reduced threshold currents for injection lasers with carrier capture into QW. To illustrate this idea, we calculate the recombination rate at the stimulated emission (SE) threshold $R_{\rm th}$ and convert it into threshold radiation intensity $I_{\rm th} = \hbar \omega R_{\rm th}/\alpha_{\rm abs}$ and threshold current $J_{\rm th} = e R_{\rm th}/\alpha_{\rm cap}$ (here $\alpha_{\rm abs}$ is the absorption probability of pump photons and $\alpha_{\rm cap}$ is the capture probability of electrons into QW which we assume to be $100\%$). The resulting threshold pumping vs lasing frequency is shown in \autoref{fig:current-frequency}, the QW thickness $d$ in this plot being swept as a parameter. 

At 77 K, the calculated threshold currents of HgCdTe QW laser diodes are around two orders of magnitude below the threshold currents of existing QCLs operating in the 10--30 $\mu$m range~\cite{QCL_17.8um_77K,QCL_24.5um_77K,QCL_28.3um_77K} due to the absence of fast phonon recombination. Beyond 15 $\mu$m, HgCdTe QW lasers are expected to outperform interband cascade lasers (ICLs) as well because of Auger suppression. However, the longest wavelength achieved by ICLs is 10.4 $\mu$m~\cite{ICL_10.4um_77K}, where HgCdTe QWs are equivalent to parabolic-band materials in terms of threshold energy and current.

Lower optical phonon frequencies of HgCdTe allow lasing within the reststrahlen band of III-V materials ($\sim$5--10 THz), where no QCLs are available. The minimum lasing frequency is achieved in near-critical QW $d=6.5$~nm ($d_c = 7.4$~nm at 77 K) and could reach 5 THz in the absence of lattice absorption. The latter is not included in our model as its effect depends on the details of laser design (number of QWs, mode confinement, etc.) but we estimate that lasing down to $\sim 6$~THz is feasible.

At room temperature, the increased intraband and intersubband absorption shifts the minimum lasing frequency to higher values, while the small energy extent of the Dirac part of the spectrum is no longer sufficient to suppress AR in narrow-gap wells. As a result, QCLs become superior in the $\gtrsim 6$--7 $\mu$m range despite the fast phonon recombination therein. The reasons are the almost empty lower lasing level in QCLs and higher joint density of states for transitions between parallel subbands, leading to lower threshold concentrations in QCLs~\cite{QCL_threshold_concentration} as opposed to interband lasers~\cite{ICL_3.67um_300K}. However, in the mid-infrared ($\lesssim6$--7 $\mu$m) HgCdTe QW laser diodes are on par with QCLs and interband cascade lasers in terms of room-temperature threshold currents while being simpler in design and fabrication.

%% file: discussion.tex
\hypertarget{discussion}{\section{Discussion and conclusions}}

 A detailed comparison of the developed theory with experiments on stimulated emission in optically pumped HgCdTe QWs~\cite{experiment,experiment-old} requires the knowledge of non-equilibrium carrier temperature. The latter is dictated by the balance of optical and Auger heating and phonon cooling. At the same time, it is possible to compare the experimentally obtained data on threshold intensity vs lasing frequency~\cite{experiment} with theory and extract the carrier temperature thereof. Such a comparison is presented in \autoref{fig:current-frequency} where experimental data for various QWs are shown with green dots. It indicates that interband pumping heats the carriers up to $\sim 80$~K, which is a reasonable value of temperature.

The presented theory shows good prospects of HgCdTe QWs for interband semiconductor lasers, in comparison with existing QCLs and ICLs. The limitations to the operating temperatures and drive currents of such lasers are still set by Auger recombination. Howsoever suppressed it was, carriers from non-Dirac parts of dispersion $\varepsilon_p$ participate in AR. This is especially pronounced for holes from the flattened parts of dispersion away from the $\Gamma$ point.

Further optimization of QW composition aiming to improve the ``diracness'' of carrier dispersion can be performed. We note, however, that increasing the fraction of cadmium \emph{in QWs} raises the energy of side maxima in the valence band, leading to a nearly thresholdless CHHH process. Reduction of cadmium fraction \emph{in the barriers} lowers the side maxima of valence band but ``spoils the diracness''. Increase in cadmium fraction in the barriers slightly increases the Auger threshold energy, but leads to denser subbands and stronger absorption.

Some prolongation of Auger lifetime can also be achieved by hole doping of QWs. A close inspection of carrier distributions at lasing threshold (see \suppref{fig:all-in-one}) shows that laser gain is indeed limited by hole population. Unfortunately, the doping-induced increase in lifetime is limited to tens of percent for QWs with $d\sim 4$~nm. 

As a result, the derived limits to Auger lifetime in HgCdTe QWs are close to fundamental ones. In wide QWs, Auger processes involving higher subbands (neglected here) may become very aggressive. However, already a fast thresholdless CHHH process shows no-go of wide (topological) wells for infrared lasers, contrary to narrow non-topological ones.

Besides Auger recombination, another serious problem for narrow-gap interband lasers can be recombination through emission of plasmons. This mechanism becomes important at high enough carrier concentrations, when the plasmon energy exceeds the bandgap, resulting in a sharp rise of threshold current in PbSnSe lasers~\cite{LeadSalt_Plasmon} and decrease in photoconductivity of bulk HgCdTe~\cite{HgCdTe_plasmon}. However, the absence of high residual doping in HgCdTe QWs in contrast to lead salts, as well as the square-root dispersion of 2D plasmons, result in plasmon energies lying below the region of interband transitions in non-topological wells~\cite{our_plasmon} (except for very narrow gaps within the HgCdTe reststrahlen band). The 2D nature of these wells also allows to additionally lower the plasmon energies by adding a metal gate, or even to benefit from plasmons by making a spaser.

To conclude, we have developed a theory of electron-hole recombination and laser gain in HgCdTe QWs and derived the fundamental limits to lasing in such structures. We have shown the possible operation of HgCdTe lasers up to $\sim 50 \mu$m wavelength at liquid nitrogen temperature. This wavelength lies within the reststrahlen band of GaAs and cannot be covered by existing GaAs QCLs.

%% file: methods.tex
\section{Methods}
 
The basic building block for calculations of recombination and optical gain are spectra and wave functions of electrons in quantum wells. We start from the eight-band anisotropic Kane model for bulk \ce{Cd_{$x$}Hg_{$1-x$}Te} with inclusion of strain effects, as described in \Cite{Kane_model}\bibnote{The sign before $\frac{18}{25}k_y\{\gamma_2-\gamma_3,k_z\}$ in the definition of $V$ should be plus, $+\gamma_2 k_x^2$ should be omitted from the definition of $R$, and $\{\kappa,k_z\}$ should be replaced with $[\kappa,k_z]$ everywhere in the definitions of $\bar{S}_{\pm}$ and $\tilde{S}_{\pm}$.}. When considering strain effects, we assumed that the QWs were grown on a thick CdTe buffer, and the lattice constants of strained barrier/well layers equal the lattice constant of unstrained CdTe. Bandstructures and wavefunctions of HgCdTe QWs were calculated using the envelope-function approximation~\cite{Kane_model} and expanding the envelope functions in a plane-wave basis. This approach has been verified experimentally~\cite{Kane_model} by comparing the calculated Landau level energies with the magnetospectroscopic data.
 
 Threshold concentrations were found from the gain condition for in-plane polarization, $(\sigma_{xx}+\sigma_{yy})/2<0$ at some frequency. (Because of rather small anisotropy at relevant electron momenta, the optical conductivities for $x$- and $y$-polarization are similar, and we used their average.) We considered all intra- and inter(sub)band contributions to the optical conductivity. Intraband contributions were taken in Drude form with constant scattering rate $\gamma = 1$~meV. This scattering rate and the corresponding spectrum broadening were also included in the inter(sub)band contributions by convolving them with a Lorentzian of width $\gamma$.

 Auger recombination rates were obtained from Fermi's golden rule by Monte Carlo integration (see \hyperlink{supplementary}{Supporting information} for details). Lattice screening was taken into account by using a frequency-dependent dielectric function in the Lorentz multi-oscillator form with the parameters taken from Table 1 of \Cite{phonon_params} (the high-frequency dielectric constant was taken to be $\kappa_{\infty} = 12$~\cite{high-frequency_kappa}; dependence of the dielectric function on alloy composition, temperature, and lattice defects was ignored).
 
 We assumed Fr\"{o}hlich electron-phonon interaction and calculated the phonon-assisted recombination rates $R_{\rm ph}$ from the following formula, which absorbs phonon frequencies and Fr\"{o}hlich coupling constants into the inverse dielectric function $\kappa^{-1}(\omega)$ and generalizes the Fermi's golden rule by taking into account finite phonon lifetimes~\cite{our}:
 \begin{equation*}
     \begin{aligned}
       R_{\rm ph} &= 2 \pi \sum_{i \in c, f \in v} (f_i \bar{f}_f \bar{n}_{\rm ph} - \bar{f}_i f_f n_{\rm ph})\\
       &\times \int_{-\infty}^{+\infty} d\omega |M_{if}(\omega)|^2 \delta(E_i-E_f-\hbar\omega),\\ |M_{if}(\omega)|^2 &= - \frac{1}{\pi} \operatorname{Im} \kappa^{-1}(\omega) \int_{-\infty}^{+\infty} dz dz' u_{if}(z)u_{fi}(z')\\
       &\times V(q_{\parallel},z-z'),
     \end{aligned}
   \end{equation*}
   where $i$ and $f$ are the initial and final electron states belonging to the conduction and valence band, respectively, with energies $E_{i/f}$ and  occupation numbers $f_{i/f}$, $n_{ph}$ denote phonon occupation numbers, $\bar{f}_{i/f} = 1 - f_{i/f}$, $\bar{n}_{\rm ph} = n_{\rm ph} + 1$,
   $u_{if} = \int d^2\vec{r}_{\parallel} \psi_{f}^{\dagger}(\vec{r}_{\parallel},z)e^{-i\vec{q}_{\parallel}\vec{r}_{\parallel}}\psi_{i}(\vec{r}_{\parallel},z)$ is the overlap between the initial and final electron states, $V(q_{\parallel},z-z') = (2\pi e^2 / q_{\parallel})\exp(-q_{\parallel}|z-z'|)$ is the Coulomb potential, $q_{\parallel}$ is the transferred wavevector, and $z,z'$ are coordinates along the growth direction.
   
 Radiative recombination rate was calculated by integrating interband conductivity $\sigma_{cv}(\omega)$ with Bose-Einstein distributions~\cite{HgCdTe_Recombination}:
 \begin{equation*}
     \begin{aligned}
       R_{\rm rad} &= \frac{8}{3\pi}\frac{\sqrt{\kappa_\infty}}{c^3}\int_0^{+\infty}\omega^2 d\omega\left[n_B(\omega-\Delta\mu_{cv})-n_B(\omega)\right]\\
       &\times \Re\frac{\sigma^{xx}_{cv}(\omega)+\sigma^{yy}_{cv}(\omega)+\sigma^{zz}_{cv}(\omega)}{2}.
     \end{aligned}
   \end{equation*}
  
Recombination processes were considered only between the lowest conduction and the upper valence subband. Electron/hole distributions were taken in the Fermi-Dirac form with temperature $T_e$ and chemical potentials $\mu_{c/v}$ for all conduction/valence subbands. Phonons were assumed to be in thermal equilibrium at the lattice temperature $T_{\rm lattice}$. QWs were assumed to be undoped.

%% file: suppdescr.tex
\begin{supplementarydescription}
Detailed description of numerical procedure for Auger recombination rate evaluation; supplementary figures showing (1) bandstructures, threshold Auger processes, and carrier distributions at threshold; (2) different optical transitions and their contributions to optical conductivity; (3) threshold concentrations and recombination times with temperature-dependent bandstructures; (4) bandgaps, lasing frequencies and threshold currents vs thickness; (5) Auger threshold energies divided into separate electron and hole contributions; (6) suppression of AR rate due to electron wave function delocalization across QW.
\end{supplementarydescription}

%% file: acknowledgements.tex
\section*{Acknowledgements}
This work was supported by the grant 16-19-10557 of the Russian Science Foundation.

%% file: main.bbl
\providecommand{\latin}[1]{#1}
\providecommand*\mcitethebibliography{\thebibliography}
\csname @ifundefined\endcsname{endmcitethebibliography}
  {\let\endmcitethebibliography\endthebibliography}{}

%% file: supporting.tex
\begin{supplementary}
\section{Calculation of Auger recombination rate}

We start from the following expression for the rate of CCCH Auger recombination (and a similar expression for CHHH recombination; the reverse processes of impact ionization are easily included by multiplying the result by $1-\exp\left[-(\mu_c-\mu_v)/kT\right]$):
\begin{eq}{Auger_rate}
       R_{\rm CCCH} = \frac{2 \pi}{\hbar} \int \frac{d^2 \vec{k}_{1,2,3}}{(2\pi)^6} |M_{fi}|^2 f_1 f_2 \bar{f}_3 (1-f_4) \delta(E_1+E_2+\bar{E}_3-E_4),
\end{eq}
($M_{fi}$ is the spin-summed Auger matrix element; $k_i$, $E_i$, $f_i$ are the momenta, energies, and occupancies of the initial (1,2,3) and final (4) states; bars indicate hole energies/occupancies).

To efficiently compute this multidimensional integral, we first rewrite the integral in new variables $E_{1,2,4}$ and $\varphi_{1,2,4}$ (polar angles corresponding to momenta $k_{1,2,4}$):
\begin{eq}{to_Ephi}
       \int d^2 \vec{k}_{1,2,3} \rightarrow \frac{1}{\hbar^3}\int\displaylimits_{\substack{
       E_4 - E_c \geq E_{\rm th}^{\rm CCCH}+E_g\\
       E_c \leq E_2 \leq E_4 - E_c - E_v\\
       E_v \leq E_1 \leq E_4 - E_2 - E_c
       }} \frac{k_{1,2,4}^2}{|\vec{v}\cdot\vec{k}|_{1,2,4}}dE_{1,2,4}d\varphi_{1,2,4},
\end{eq}
because the sharpest parts of the integrand ($\delta$-function and Fermi-Dirac distributions) depend only on energies. The integration region is bounded by the conduction/valence band edges $E_{c/v}$ and the threshold energy of CCCH (CHHH) process $E_{\rm th}^{\rm CCCH}$ ($E_{\rm th}^{\rm CHHH}$).

Then, we remove the $\delta$-function by integrating over $\varphi_2$:
\begin{eq}{removing_delta}
       \int \frac{k_2^2}{|\vec{v}_2\cdot\vec{k}_2|}dE_{1,2,4}d\varphi_{1,2,4}\delta(E_1+E_2+\bar{E}_3-E_4) \rightarrow \frac{1}{\hbar}\int \frac{dE_{1,2,4}d\varphi_{1,4}}{|\vec{v}_2 \times \vec{v}_3|}.
\end{eq}
The integrand still depends on $\varphi_2$, which we cannot express through other variables analytically, so at each integration point we find $\varphi_2$ by solving the momentum and energy conservation numerically.

The remaining five-dimensional integral is not yet well-suited for straightforward Monte Carlo integration, since Fermi-Dirac distributions are not slowly varying functions, especially in the Boltzmann limit. To make the integrand smooth, we perform importance sampling drawing samples from $(1/Z) f_1 f_2 \bar{f}_3 (1-f_4)$ distribution ($\varphi_{1,4}$ are sampled uniformly over $[0, 2\pi]$). This is accomplished by (1) sampling $E_4$ from $(1/Z')\{\exp[E_4-\mu_1-\mu_2-\mu_3]+1\}^{-1}\{\exp[\mu_4-E_4]+1\}^{-1}$ distribution (it approximates $(1/Z) f_1 f_2 \bar{f}_3 (1-f_4)$ with a univariate distribution and allows inverse transform sampling); (2) applying rejection sampling to get the desired $(1/Z) f_1 f_2 \bar{f}_3 (1-f_4)$ distribution. The normalization factor $Z$ is computed by straightforward numerical integration.

Finally, there remains the problem of infinite variance: after changing variables and removing the $\delta$-function, the integrand acquires a denominator which can become zero at some points, so that the integrand has a finite mean, but infinite variance. This breaks the usual argument about square-root convergence of Monte Carlo integration. Actually, the Monte Carlo algorithm still converges~\cite{Monte_Carlo_infinite_variance}, but more slowly than in the finite-variance case. To improve the convergence and get rid of `ripply' plots, we employ the truncated mean technique widely used in statistics for this particular purpose, estimating means of heavy-tailed distributions. To put it simple, we discard one or two largest samples of the integrand, which is enough to make the distribution of the remaining samples have a finite variance, while introducing only a minor systematic underestimation of the integral.

\clearpage

\section{Supplementary figures}

In this section, we present some supplementary data.

\autoref{fig:all-in-one} visually represents many properties of HgCdTe QWs, such as bandstructure, threshold Auger processes, threshold concentration together with the corresponding quasi-Fermi levels, lasing wavelength and threshold current, and their evolution with QW thickness and temperature.

\autoref{fig:optical_conductivity} illustrates the complex interplay between interband gain and intersubband absorption in narrow-gap HgCdTe QWs by showing the possible optical transitions together with their ``strengths'' (with joint density of states also taken into account).

Threshold concentrations and recombination times are shown in \autoref{fig:n_and_t} in the case of equal lattice and electron temperature (this extends Figures \ref{fig:concentrations}, \ref{fig:recombination_times} of the main text, where lattice was held at 4.2 K).

\autoref{fig:currents_and_gaps} supplements \autoref{fig:current-frequency} of the main text by showing lasing frequencies and threshold currents vs QW thickness. Bandgaps and optical gaps are also given for comparison.

\autoref{fig:e-h_thresholds} shows the Auger threshold energies divided into contributions from each involved carrier. One can see that the threshold energy is determined mostly by the hole energies. This means that hole doping would substantially increase the recombination rate, negating the benefit of more balanced electron/hole populations.

Finally, the effect of Auger recombination suppression in HgCdTe QWs in comparison with truly 2D materials due to the finite extent of electron wavefunctions in the direction perpendicular to the well is shown in \autoref{fig:3Dvs2D}. The Auger suppression factor is defined as the ratio between the Auger recombination time calculated with 3D Coulomb interaction $V(q_{\parallel},z-z') = (2\pi e^2 / q_{\parallel})\exp(-q_{\parallel}|z-z'|)$ and the same time calculated with 2D Coulomb interaction $V(q_{\parallel},z-z') = 2\pi e^2 / q_{\parallel}$. An approximate expression for this factor can be obtained by taking $\exp(2 q_{\parallel}|z-z'|)$ (because Coulomb interaction is squared in the Auger matrix element) with $q_{\parallel}$ substituted with $q_{\rm th}^{\rm CCCH}$, the momentum transfer in the threshold CCCH process, and $|z-z'|$ substituted with $d/2$, half the QW thickness.

\begin{widefig}{all-in-one}{figures/Supplementary/All-in-one}
Calculated bandstructures, threshold CCCH and CHHH (if allowed by conservation laws) Auger processes, electron (red) and hole (cyan) distributions at lasing threshold, lasing transitions (orange), and threshold currents at liquid nitrogen and room temperature for \ce{Cd_{0.7}Hg_{0.3}Te/HgTe/Cd_{0.7}Hg_{0.3}Te} QWs of different thicknesses grown along the [013] direction. Red and cyan stripes indicate the quasi-Fermi levels and thermal energy (more precisely, they depict regions $\mu_c \leq E \leq\mu_c + kT$ and $\mu_v - kT \leq E \leq \mu_v$). Band anisotropy is shown by the thickness of blue curves (artificially increased by 4 meV for legibility).
\end{widefig}

\begin{widefig}{optical_conductivity}{figures/Supplementary/Optical_conductivity} Possible vertical inter(sub)band transitions in a 6 nm \ce{Cd_{0.7}Hg_{0.3}Te/HgTe/Cd_{0.7}Hg_{0.3}Te} well at 77 K and their contributions to the optical conductivity without taking into account the populations of lower and upper state. More precisely, the plotted quantity $C_{\alpha\beta}(k,\omega)$ is proportional to the oscillator strength and joint density of states and yields optical conductivity after multiplying by carrier distributions and averaging over the directions of the electron momentum: $\sigma_{\alpha\beta}(\omega) = \int_0^{2\pi}\frac{d\varphi}{2\pi}\sum_{ss',k:E_s(k,\varphi)-E_{s'}(k,\varphi)=\hbar\omega}
\left[ f_{s'}(k,\varphi)-f_s(k,\varphi) \right] C_{\alpha\beta}(k,\omega)$. Left plot: contributions $(C_{xx}(k,\omega)+C_{yy}(k,\omega))/2$ to the in-plane optical conductivity, right plot: contributions $C_{zz}(k,\omega)$ to the out-of-plane conductivity. Subbands are ordered by the distance from the midgap: e.g., ``c1-v1'' means transitions between the lowest conduction and highest valence subband.  Regions of high $C_{\alpha\beta}(k,\omega)$ (white, yellow) appear at the singularities of the joint density of states. The minimum thickness of the curves is artificially limited to 1 THz for legibility.
\end{widefig}

\begin{widefig}{n_and_t}{figures/Supplementary/Concentrations_and_times}
The same as Figures~\ref{fig:concentrations}, \ref{fig:recombination_times} in the main text, but with lattice temperature equal to the electron temperature.
(a) Threshol carrier concentrations in \ce{Cd_{0.7}Hg_{0.3}Te/HgTe/Cd_{0.7}Hg_{0.3}Te} QWs vs thickness and temperature. Solid curves: $\gamma = 0$, only intersubband absorption is considered. Dashed curves: $\gamma = 1$ meV, both intersubband and Drude absorption are included. (b) Auger (solid curves), phonon (dashed curves), and radiative (dotted curves) recombination times in \ce{Cd_{0.7}Hg_{0.3}Te/HgTe/Cd_{0.7}Hg_{0.3}Te} QWs of different thicknesses at threshold carrier density. The counterintuitive increase of phonon recombination time with temperature is due to (1) wider bandgap, (2) increased population of high-energy states that cannot take part in phonon recombination. Error bars show the standard deviations coming from Monte Carlo integration.
\end{widefig}

\begin{widefig}{currents_and_gaps}{figures/Supplementary/Currents_and_gaps} (a) Bandgaps (solid curves), optical gaps (dashed curves), and lasing frequencies at threshold (dash-dotted curves) of \ce{Cd_{0.7}Hg_{0.3}Te/HgTe/Cd_{0.7}Hg_{0.3}Te} QWs of different thicknesses at 300 K (red) and 77 K (blue). Electron temperature equals lattice temperature. (b) Threshold currents vs thickness of \ce{Cd_{0.7}Hg_{0.3}Te/HgTe/Cd_{0.7}Hg_{0.3}Te} QWs at 300 K (red) and 77 K (blue). Threshold intensities of optical pumping are shown on the right axis, assuming 0.4\% absorption of a 2.15 $\mu$m pump, as in \Cite{supp_experiment}. Solid curves: lattice temperature equals electron temperature, dashed curves: lattice is held at 4.2 K.
\end{widefig}

\begin{figure*}
\centering
  \begin{halffig}{e-h_thresholds}{figures/Supplementary/e-h_thresholds}
  Energies of the carriers involved in threshold Auger processes. Threshold CCCH process involves two electrons with the same energy (red curve) and a hole (green curve). Threshold CHHH process involves two holes with the same energy (black curve) and an electron (blue curve). All the energies are relative to the band edges; they were found by a numerical algorithm with 1 meV accuracy. Lattice temperature is 4.2 K.
  \end{halffig}
\hskip 0.02\textwidth
  \begin{halffig}{3Dvs2D}{figures/Supplementary/3Dvs2DCoulomb}
  Solid curve: Auger suppression factor (defined in the text) due to the non-truly-2D nature of \ce{Cd_{0.7}Hg_{0.3}Te/HgTe/Cd_{0.7}Hg_{0.3}Te} QWs. Dashed curve: Auger suppression factor calculated using an approximate expression $\tau_{\rm Auger}/\tau_{\rm Auger}^{\rm 2D} \approx \exp(q_{\rm th}^{\rm CCCH}) d$. Lattice and electron temperature is 77 K.
  \end{halffig}
\end{figure*}
\clearpage
\end{supplementary}

%% file: supporting.bbl
\renewcommand{\refname}{Supplementary references}
\providecommand{\latin}[1]{#1}
\providecommand*\mcitethebibliography{\thebibliography}
\csname @ifundefined\endcsname{endmcitethebibliography}
  {\let\endmcitethebibliography\endthebibliography}{}

%% file: main.bbl
\begin{mcitethebibliography}{54}
\bibitem[ani()]{anisotropynote}
Actually, the side maximum of the valence band is circular only in the
  isotropic approximation. When the band anisotropy present in HgCdTe QWs is
  taken into account (which we did, see \suppref{fig:all-in-one}), the valence band edge consists of only two points
  in the momentum space. However, this still results in a zero-threshold Auger
  process along at least some directions in the $k$-space, and in elevated
  density of states near the valence band edge. In view of the above, we will
  call this maximum `circular' for simplicity.\relax
\mciteBstWouldAddEndPunctfalse
\mciteSetBstMidEndSepPunct{\mcitedefaultmidpunct}
{}{\mcitedefaultseppunct}\relax
\EndOfBibitem
\bibitem[lea()]{leadsaltnote}
In \Cite{LeadSalt_DoubleHeterostructure}, the lasing wavelength is $\lambda$ =
  20 $\mu$m at 20 K. We assume it decreases to 15 $\mu$m at 77 K based on the
  known temperature dependence of $\lambda$ in
  \ce{Pb_{$1-x$}Sn_{$x$}Se}~\cite{Lead_Salt_LaserCharacteristics}.\relax
\mciteBstWouldAddEndPunctfalse
\mciteSetBstMidEndSepPunct{\mcitedefaultmidpunct}
{}{\mcitedefaultseppunct}\relax
\EndOfBibitem
\bibitem[Not()]{Note-1}
The sign before $\frac{18}{25}k_y\{\gamma_2-\gamma_3,k_z\}$ in the definition
  of $V$ should be plus, $+\gamma_2 k_x^2$ should be omitted from the
  definition of $R$, and $\{\kappa,k_z\}$ should be replaced with
  $[\kappa,k_z]$ everywhere in the definitions of $\bar{S}_{\pm}$ and
  $\tilde{S}_{\pm}$.\relax
\mciteBstWouldAddEndPunctfalse
\mciteSetBstMidEndSepPunct{\mcitedefaultmidpunct}
{}{\mcitedefaultseppunct}\relax
\EndOfBibitem
\end{mcitethebibliography}
